# A comparative study on the bottleneck flow between preschool children and adults under different movement motivations

Hongliu Li[1,3], Jun Zhang[1*], Libing Yang[2], Weiguo Song[1], Kwok Kit Richard Yuen[3]

1 State Key Laboratory of Fire Science, University of Science and Technology of China, Jinzhai Road 96, Hefei, Anhui, People's Republic of China
2 Jingzhou Municipal Bureau of Tobacco Monopoly, Jingzhou Hubei 434000, People's Republic of China
3 Department of Architectural and Civil Engineering, City University of Hong Kong, Kowloon 999077, Hong Kong, People's Republic of China

## Highlights

- We analyze bottleneck flow of preschool children in a laboratory experiment.
- The pedestrian flow of children and adult under different movement motivations are compared.
- The relation between flow and bottleneck width for children and adults are unified.
- Arch-like and teardrop-shape distribution are observed for the children' and adults' movement respectively.

## Abstract:

Understanding on children's movement characteristics is significant to improve their safety levels especially under emergencies. In this work, we performed laboratory experiment to investigate the dynamics of preschool children passing through bottlenecks under high movement motivation. It is found that the relations between flow and bottleneck width for adults and children can be unified when the body size and movement motivation are considered in certain way. High movement motivation leads to competition among pedestrians, which results in different spatial-temporal distribution of the density and speed compared to normal movement of adult. Based on Voronoi method, an arch-like distribution is observed for the children movement with competition, whereas it is teardrop-shaped for the adults without competition. The peak density of children, which reaches nearly 14 ped/m$^2$, appears at the place 30 cm away from the bottleneck entrance. However, it is about 100 cm to the bottleneck for adults. Besides, several typical behaviors of children like guiding, pushing as well as playing are observed in the experiment. The findings in this study will benefit the evacuation drill design as well as the facility design for children.

---

* Corresponding author：junz@ustc.edu.cn

# 1. Introduction

According to World Population Prospect (Nations, 2017), the children under 9 years represents about 17.9% of the world's population. Among them the ones between 3 to 5 years are a special group, since most of them stay in kindergarten for a long time according to National Center for Education Statistics (NCES, 2017) but with weaker self-care ability compared to the pupils. Children aged 3-5 years old are learning how to control their bodies. They easily lose their balance and fall down. Besides, small children are easily to be distracted by other things and fail to follow a desired direction. Their unmatured physical and psychological characteristics make them more vulnerable and facing with higher risks both in daily life and in emergency (Kholshevnikov et al., 2009). Furthermore, more children are enrolled in preschool program nowadays and they have to stay in kindergartens with their teachers and peers. It is unrealistic to ensure that children are under supervision all the time and they have to move in groups when facing emergency. To make reasonable evacuation plans and comfortable buildings for children, it is necessary to pay attention to children's safety and to understand their movement characteristics.

Bottleneck is a common geometry in most of pedestrian facilities. When a large groups of discrete bodies like pedestrians, small steel balls and sheep pass through a narrow exit, it is foreseeable that they might accumulate in front of the exit since the narrowing restrict their movement (Rupprecht et al., 2011). Due to the flow limitation of the bottleneck, stampede disasters often occur near the bottleneck (Helbing et al., 2007; Wikipedia, 2014). These serious disasters have taken away lives of many people and get a number of people injured. The serious consequences arouse attention of many researchers and many studies have been performed focusing on the bottleneck flow of pedestrian in the past two decades (Daamen and Hoogendoorn, 2010; Helbing et al., 2000; Helbing et al., 2007; Rupprecht et al., 2011; Sun et al., 2017; Zuriguel et al., 2014). It is pointed out that pedestrian movement through a bottleneck is influenced by several factors including the geometry of the bottleneck (length, width and location) (Kretz et al., 2006; Liddle et al., 2009; Sun et al., 2017), the initial distribution of pedestrians, population composition, light intensity as well as sociological effects (Daamen and Hoogendoorn, 2010; Liddle et al., 2009). The impact of bottleneck lengths, widths and shapes on the flow has been investigated widely with different objects such as pedestrians, sheep herds or mice (Liddle et al., 2009; Sun et al., 2017; Zuriguel et al., 2014). It is found that an appropriate shape can improve the traffic capacity (Sun et al., 2017). Pedestrian's flow shows an upward trend with the increase of the bottleneck width and it oscillates around a value when the bottleneck width is large (Kretz et al., 2006; Liddle et al., 2009; Seyfried et al., 2009). Several scientists put forward that the peak density area appears before the bottleneck and the congestion is impacted by the bottleneck width (Liao et al., 2014; Liddle et al., 2011a). The process of the lane formation is affected by the bottleneck width (Liddle et al., 2009; Liu et al., 2014; Seyfried et al., 2009) and the distribution of time gaps between two consecutive pedestrians are found to be symmetrical with distinct maximum at the center and becomes flatter with the increase of the bottleneck width (Kretz et al., 2006). The probability of occurrence of time gap is consistent with the existence of a power-law tail (Lin et al., 2016; Pastor et al., 2015). Zipper effect which increases the capacity of the bottleneck is identified from experiments (Hoogendoorn and Daamen, 2005). Paradoxical phenomena like faster-is-slower effect can be observed from experiments and simulations (Garcimartin et al., 2016; Helbing et al., 2000; Pastor et al., 2015). The performance of an obstacle near an exit on pedestrian flow is summarized and no specific relation between the bottleneck width, the obstacle distance and obstacle size/shape is found. (Shiwakoti et al., 2019) Even though there are a number of studies focused on flow through bottlenecks, most of them obtain data from animals, particles or adults. Up to now, very few studies on the movement characteristics of children can be found, especially for preschool children (Kholshevnikov et al., 2009; Larusdottir and Dederichs, 2011). Rosenbloom et al. study road crossing behavior of children aged 6 to 13 and put forward that children perceive the road crossing behavior of their peers is riskier than that of their own and they might run into the road without checking for traffic at first (Rosenbloom et al., 2012). Some researchers observe that the use of handrails or walking hand in hand can improve the speed of children aged 3-5 on the stairs (Larusdottir and Dederichs, 2012). Children own higher walking speed in spiral stairs and lower speed on horizontal plane (Larusdottir and Dederichs, 2012). Children aged 5-6 move faster than adults at the same density and the evacuation process of children is influenced by the guidance of teachers, stairs and congestion on the transition platform (Fang et al., 2019). Compared to adults, the density and flow of children are higher when they pass the door (Kholshevnikov et al., 2009). The strength capabilities of children aged 6 is evaluated and it is pointed out that children are unable to operate the rolled-over school bus rear emergency door of maximum designed force and the last row of seats impact the flow of evacuation children. (Abulhassan et al., 2018) However, there is still a lack of quantitative description of children movement characteristics up to now. Researches on the dynamics of children movement needs to be supported by richer data and the difference between children and adults is still lack of in-depth study. The lack of data and special behavior characteristics of children determine the necessity and urgency to study children's evacuation characteristics through bottlenecks.

In this study, preschool children experiments are performed to investigate the impact of the bottleneck width on children's evacuation. We study the relationship between flow and the bottleneck width, density and speed distribution of children before the bottleneck, and make comparison between children and adults. The findings in this study will benefit the evacuation design for children and promote to carry out evacuation drills in kindergartens to improve safety levels of children. The remainder of the paper is as following. In section 2, the experiment setup is briefly described. Section 3 shows the main results from the experiment analysis and the concluding remarks are made in section 4.

## 2. Experiment Setup

The experiment was performed in March 2018 for a day at a kindergarten in Yueyang, Hunan province, China. The participants were 54 preschool children (24 girls and 30 boys) aged 3-5 in the kindergarten. Their mean height and weight were 1.07 m and 18.27 kg, respectively. Fig. 1 shows a snapshot of the experiment. The boundary of the scenario was built with security fences (height: 0.75 m) and boards (size: 1.20 m × 1.00 m × 0.05 m). The edges of boards were covered by foam material to avoid bruising during the experiment. To improve the stability of boards, stones as well as managing staffs have played a supporting role. The width of exit was changed from 0.4 m to 1.1 m at 0.10 m intervals to study their influence on the flow. Initially, the children stood within the "Waiting Area", which was 5.00 m away from the entrance of the bottleneck and can provides adequate longitudinal space for children to accelerate. The initial density ρ is about 4 ped/m$^2$. When the children were ready to start, we gave them a command "Ready! Go!" and the experiment started. The children were very high motivated during the experiment and tried to leave the bottleneck as fast as they can. Considering this phenomenon, we no longer tried to control their movements or ask them to move in normal situation since they were too young to understand complex instructions. As a result, we only told them to pass through the bottleneck as fast as possible and carefully.

The experiment was performed in a whole morning for the restriction of children' weak physical strength and limited time. A total of 26 runs were performed and each scenario performed more than twice if there were children falling down during the experiment. Since the children' limited physical strength, the number of children varied in each run as shown in Table 1. In this study, we only use one run for each width in which nobody fell down.

Table 1. Number of runs in the total experiment and number of children in selected experiment.

| Bottleneck width [m] | 0.40 | 0.50 | 0.60 | 0.70 | 0.80 | 0.90 | 1.00 | 1.10 |
|---|---|---|---|---|---|---|---|---|
| Number of runs | 5 | 4 | 3 | 3 | 3 | 2 | 3 | 3 |
| Number of children | 54 | 52 | 51 | 48 | 50 | 50 | 49 | 46 |

During the experiment, each kid was asked to wear a hat (blue, yellow or pink) for easy detecting and tracking afterward. One HDR-SR11 camera (resolution: 1920 x 1080 pixels, frame rate: 25 fps) located in the third floor recorded the whole process during the experiment. Before and after the experiment, we calibrated the camera and the whole scenario twice based on the operation instruction of the software *PeTrack* (Boltes et al., 2010) to extract the trajectories precisely. We extracted all the trajectories semi-automatically from video recordings of HDR-SR11 camera. Moreover, all quantitative analysis following is based on the trajectory data.

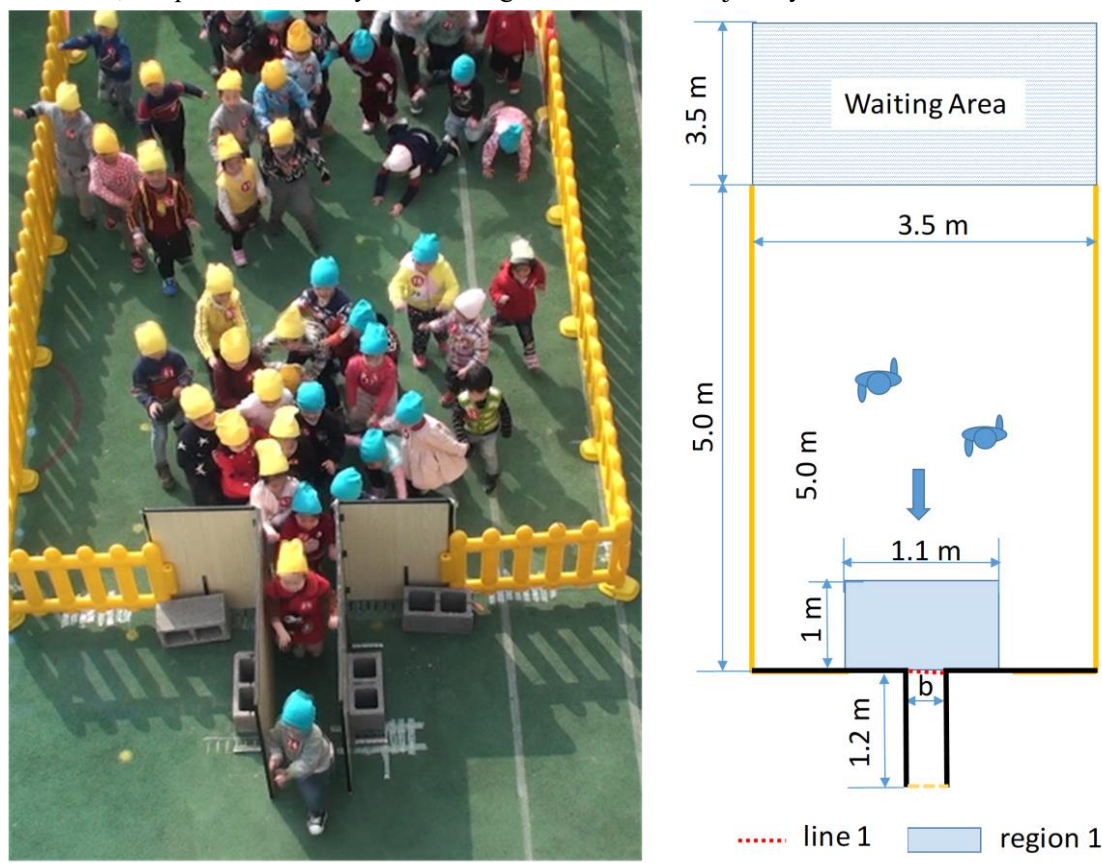

Fig. 1 A snapshot of and the illustration of the experiment. About 50 children (3-5 years old) initially stood in the waiting area and ran out of the bottleneck when the staff gave the command. The initial pedestrian density is about

4 ped/m². The width of the bottleneck $b$ changes from 0.4 m to 1.1 m at 0.1 m intervals. The analysis region is 1.1 m x 1.0 m (region 1) in front of the bottleneck.

## 3. Results and Analysis

In this section, we investigate the bottleneck effects in the experiment. All the analyses below are based on trajectories and recorded videos from the experiments.

### 3.1 Trajectories

Children run from the waiting area to the bottleneck. Since the capacity of bottleneck is insufficient for crowds, congestions occurred in front of the bottleneck. Fig. 2 shows all trajectories with instantaneous speed of the children passing through the bottlenecks of various widths. The children tried to pass through the bottleneck as quickly as possible and they competed against each other in front of the bottleneck, which resulted into chaotic trajectories around the bottleneck. Further, the passing capacity of bottleneck increases and the degree of chaos of trajectories decreased with the increase of the bottleneck width.

Lane formation is considered as a typical self-organization phenomenon and it is often pointed out in previous studies on pedestrian movement through bottlenecks (Liddle et al., 2009; Seyfried et al., 2009). However, no obvious lanes are observed in this experiment as shown in Fig. 2. That maybe resulted from different movement motivations of pedestrians during the experiments. Pedestrians in previous experiments passed through the bottleneck more politely and in normal speed without pushing (Kretz et al., 2006; Liddle et al., 2009; Seyfried et al., 2009; Sun et al., 2017), whereas children in this experiment went through the bottleneck quickly and behaved competitively. Some children even pushed other children to move through the bottleneck faster.

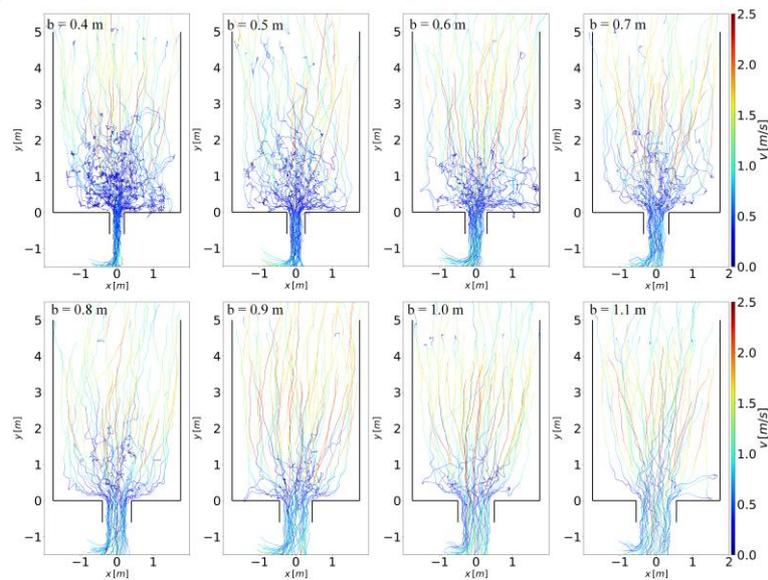

Fig. 2 Trajectories of all children in the experiment. The color indicates the instantaneous speed of each child. The trajectories are chaotic and the speeds are lower around the bottleneck entrance since children decelerate and compete against others. The degrees of chaos decrease with the increase of the bottleneck width. Children who run in front of the queue own a higher speed about 2.5 m/s.

Overall, the movement of the children in the experiment can be divided into four stages (acceleration stage, deceleration stage, congestion stage and acceleration stage). As shown in Fig. 3, the children firstly accelerated until they reached around y = 2.0 m, then they had to decelerate and went into congested area. When they passed through the bottleneck entrance y = 0.0 m, their speeds began to increase from a very low value to approximately 1.0 m/s. The limited traffic capacity of bottleneck leads to deceleration and congestion in front of the bottleneck. The impact range, the number of waiting children and the time of congestion reduce with the increase of the bottleneck width. The speed varied in a wide range for different children based on their physical characteristics and their locations in the crowds. The speeds of the children in front of the crowd can reach about 2.5 m/s to 3.0 m/s, whereas it was around 1.0 m/s for children at the end of the crowd. Further, the speed oscillates around 1.0 m/s inside the bottleneck no matter how wide the bottleneck is or how fast the children ran at beginning. Further study on the speed is discussed in section 3.2.

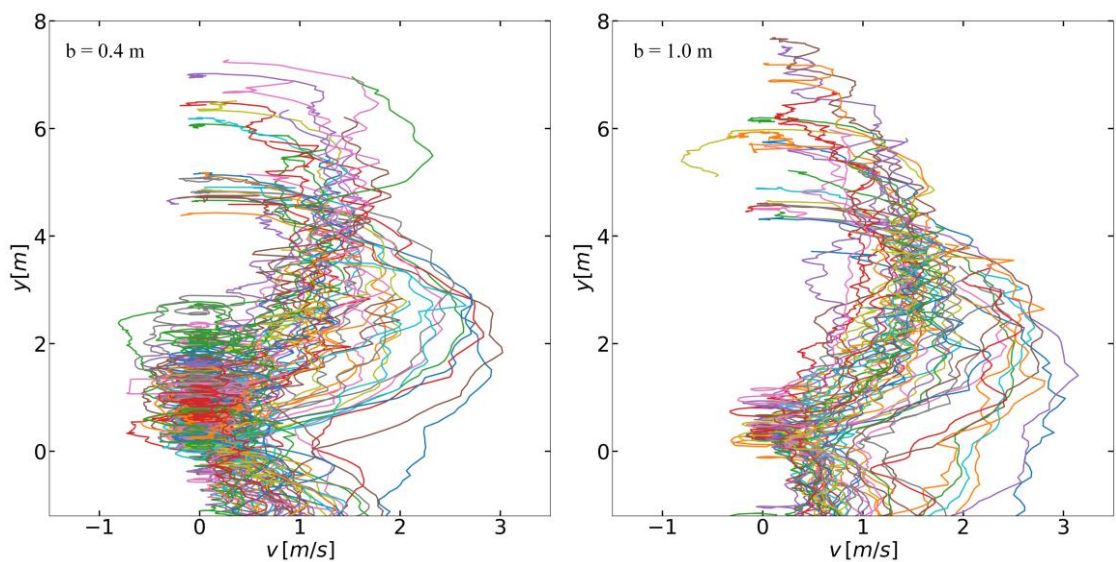

Fig. 3 The evolution of speed over space in different runs. The movement of the children can be divided into four stages: acceleration stage, deceleration stage, congestion stage and acceleration stage. The children firstly accelerated until they reached around y = 2.0 m, then they decelerated and went into congested area. After passing through the bottleneck entrance y = 0.0 m, children began to increase their speed from a very low value to approximately 1.0 m/s.

Besides, we observe several interesting phenomena among children during the experiment. For example, it is not in accordance with the general principle "first arrive, first leave". Some children behaved aggressive and even pushed others who reached the bottleneck earlier to move through the exit quickly. As shown in Fig. 4(a), the kid in the oval reached the bottleneck 3 seconds earlier than the one in the dotted ellipse. However, the latter left the exit 0.8 second earlier by pushing the former away. What's more, the pedestrian flow in front of the bottlenecks showed a stop-and-go phenomenon due to the pushing and following behaviors. The stop state and the go state alternatively occurred in the left and right areas in Fig. 4(b). The transferring between the two states is mainly dominated by the appearance of the more aggressive and stronger children at the bottleneck. Besides, we also observed behaviors like guiding, humility as well as game playing. As shown in Fig. 4(c), the kid in the oval stayed inside the bottleneck for a while and guided other children to pass the bottleneck firstly and then left. In Fig. 4(d), the kid in dotted ellipse let the one in solid ellipse go into the bottleneck firstly. The kid in rectangular in Fig. 4(d) moved out of the bottleneck slowly by holding the board, while in Fig. 4(e), the kid in solid ellipse walked slowly to block the way of the others behind him on purpose. Another point worth mentioning is that the children response differently to the same instruction and easily to be distracted by other things. From this point of view, it would be better to remove things on the evacuation route like toys that may distract children from evacuation. The uncooperative behaviors like blocking the way will reduce evacuation efficiency and increase the risk of casualty in reality. In order to improve safety level of children, it is necessary to perform evacuation drills in the kindergarten to teach children how to escape in emergency.

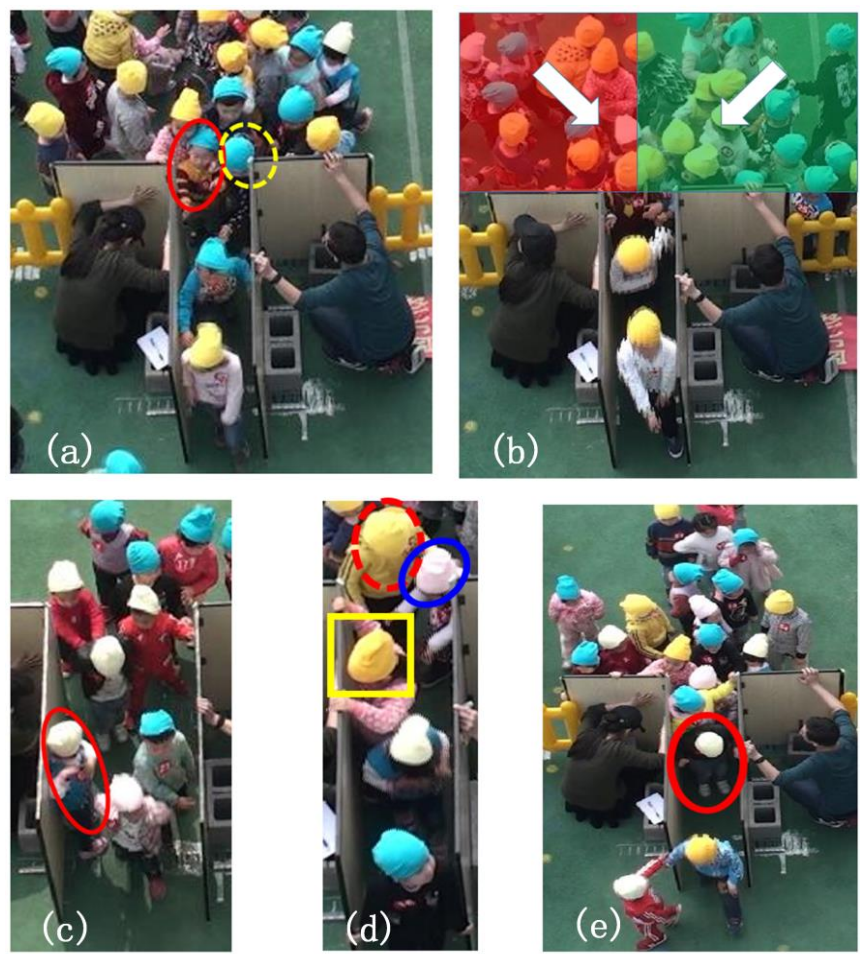

Fig. 4 Several phenomena of the children during the experiment. (a) The child in solid ellipse pushed the one in dotted ellipse away and passed the exit firstly, even though he arrived at the entrance later. (b) The stop and go state were observed in front of the bottleneck due to the competitive behaviors among children. The stop state and go state occurred alternatively in the left and right areas marked. (c) The child in the ellipse stood there and guided others to leave the bottleneck firstly. (d) The child in dotted ellipse let the one in solid ellipse leave firstly, while the child in the rectangle left the bottleneck by holding the board. (e) The child in the ellipse decreased his speed deliberately to block the way of the others behind.

## 3.2 Density and speed distribution

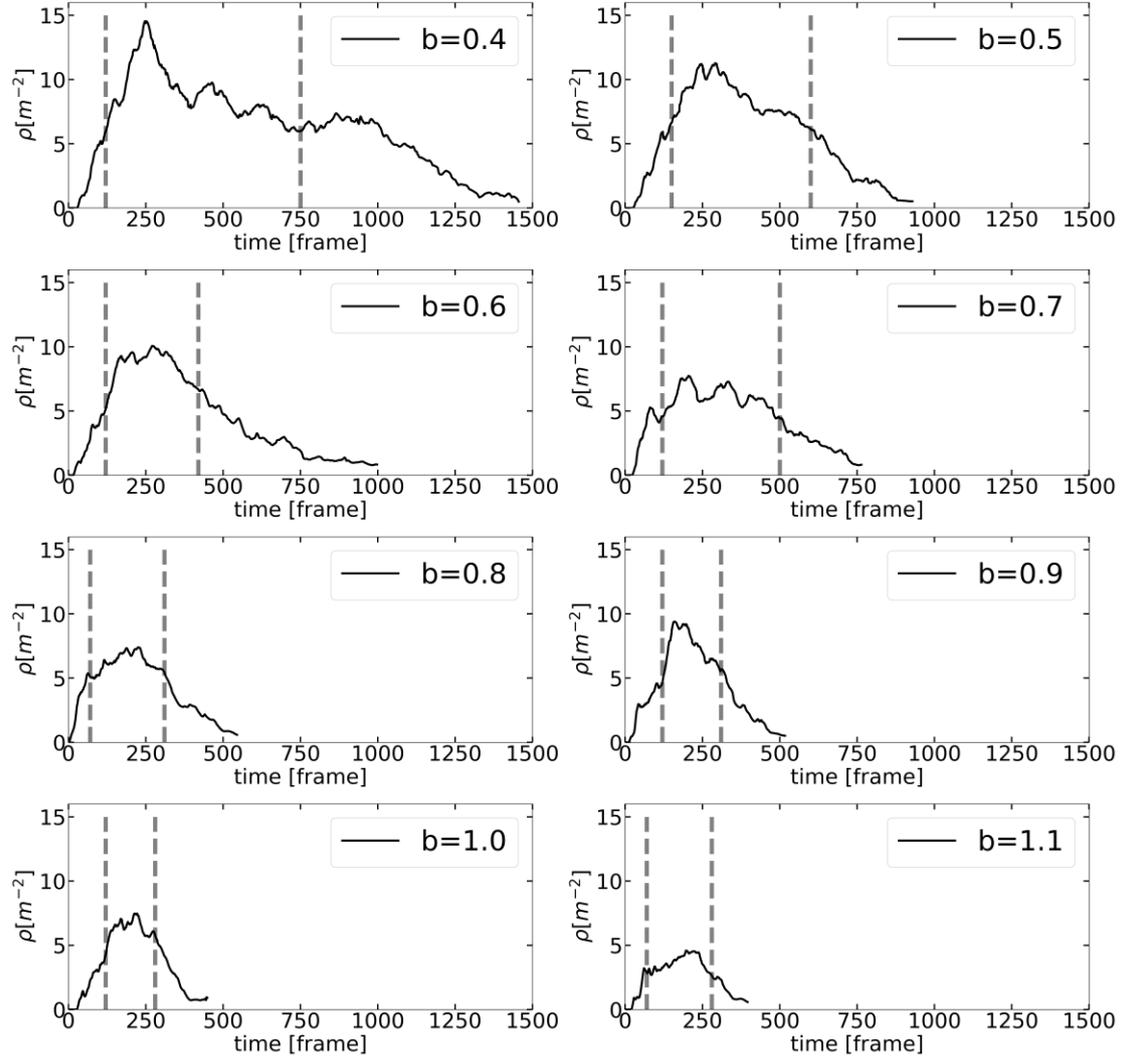

Fig. 5 Time series of density in the region 1. The vertical dashed lines indicate the start and end of the relatively steady congested state. (When *b* is 0.4 m, only data between 150 frame and 750 frame are selected since one kid obstructs exit intentionally from the 750th frame.) When b ⩽ 0.6 m, the density can be beyond 10 ped/m².

In Fig. 5, we show the time series of density in front of the bottleneck (region 1) based on Voronoi method (Steffen and Seyfried, 2010) to quantify the characteristics of pedestrians' movement. When the width is smaller than 0.6 m, the mean density can be beyond 10 ped/m², which is obviously higher than 5 ped/m² of adults in previous experiment (Zhang and Seyfried, 2014). When the width is 0.4 m, the maximum density reaches 14.5 ped/m² which is higher than 10 ped/m² in the crowd disaster in Mina during the Hajj (Helbing et al., 2007). However, considering the obvious different body size of children and adults, it is necessary to consider the dimension of human body when comparing densities. It is suggested that the maximum physiological size of human body is mainly determined by shoulder width (*w*) and chest depth (*d*) (Pheasant, 2014). As a result, we mainly consider these two parameters in the following analysis and the median values of shoulder breadth and chest depth of Chinese children and European adults are listed in Table 2. We assume the projection of the body as a rectangle and then the area *S* occupied by an individual can be calculated from $S = w \times d$. The occupied area of a kid is about 0.04 m² while it is 0.12 m² for adults. Since the required minimum area of a kid is much smaller than that of an adult, it is understandable that the density in front of the bottleneck is obviously higher than that of adults in the previous experiments (Zhang and Seyfried, 2014). If we make a conversion by considering the body size, 15 ped/m² of children actually reflects the same congestion degree with that of 5 ped/m² for adults.

Table 2. Median value of shoulder breadth and chest depth (Jürgens et al., 1998; Standardization, 2010)

| People | Shoulder breadth/m | | Chest depth/m | |
| --- | --- | --- | --- | --- |
| | Male | Female | Male | Female |
| Chinese children | 0.286 | 0.282 | 0.140 | 0.141 |
| European adults | 0.474 | | 0.215 | |

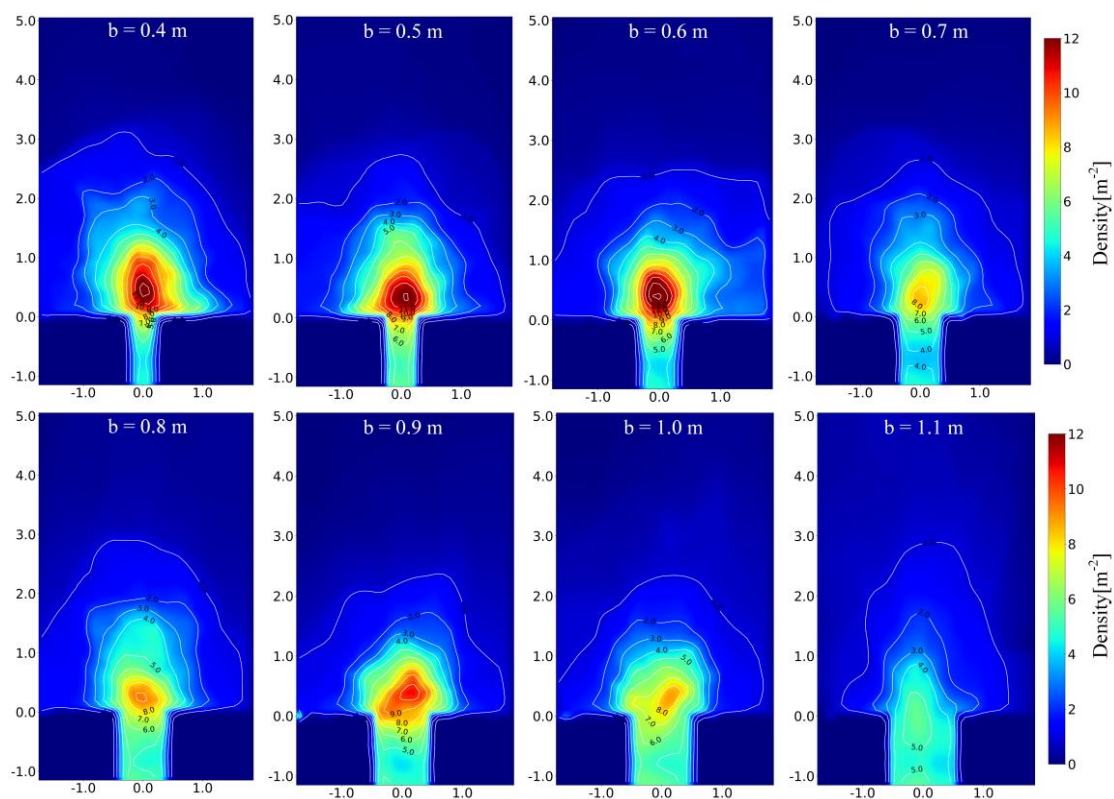

Fig. 6 Density profiles for bottlenecks of different width. The distribution of crowds in front of the bottleneck is arch-like and the shapes of peak density regions look like ellipses. The location of peak density is around 30 cm from the bottleneck entrance except b = 1.1 m. The congestion decreases with the increase of the bottleneck width.

With the Voronoi diagram based method proposed in (Steffen and Seyfried, 2010; Zhang et al., 2012), we further calculate the density and speed over small regions (10 cm x 10 cm) and obtained the maps of the observables over the experimental areas for different bottleneck widths (see Fig. 6 and Fig. 7). To reduce the fluctuations triggered by the start and the end of a run, we only consider data in the above-mentioned steady congested states to calculate these observables. To provide additional information like the location of the maximum density and directions of steepest decline, we present these maps with contours with a Gaussian filtering procedure. The Gaussian filter with sigma of 0.8 is used to filter the data of density. As described in (Liddle et al., 2011a), these maps can provide insights into the dynamics of the motion and the sensitivity of the integrated quantities to influences such as boundary geometry and measurement placement. To make comparison with results of adults easily, we plot the speed and density maps of German adult experiment in (Liddle et al., 2011a) with the same procedure (see Fig. 8).

It is clear that the distributions of density are inhomogeneous over space and transitions from high density to low density can be observed. Similar to the adult movement in (Liddle et al., 2011a), narrower bottlenecks lead to a higher density in front of the constriction. However, the congested areas in front of the bottleneck display an arch-like distribution and the shapes of peak-density regions are similar to ellipses in this study, whereas they are both teardrop-shape distributions for the adult experiment in Fig. 8(a) (Liddle et al., 2011a). When pedestrians pass the bottleneck competitively, they choose to gather around the exit and try to find a gap and push their way through the crowds. Oppositely, when pedestrians pass the bottleneck cooperatively in a walking speed, pedestrians prefer to follow the other people and wait to go out of the bottleneck from the recorded videos (Liddle et al., 2011a). These two different kinds of motivation lead to various spatial distribution of pedestrian in front of the bottleneck.

Furthermore, the same as previous studies (Liao et al., 2014; Liddle et al., 2011a; Zhang and Seyfried, 2014), the highest densities appear in front of bottleneck regardless of the bottleneck width (except b = 1.1 m). The locations of the peak density lie around 30 cm from the bottleneck entrance in this study when the bottleneck width ranges from 0.4 m to 1.0 m, whereas they are around 100 cm from the entrance of the bottleneck (b = 0.9, 1.0, 1.1 m) in former study (Liddle et al., 2011a) (see Fig. 8(a)). The value of the peak density shows a downward trend with the increase of the bottleneck width as expected. The highest density can be over 12 ped/m$^2$ for b $\leqslant$ 0.6 m and be over 8 ped/m$^2$ for b $\leqslant$ 1.0 m. When b is 1.1 m, the traffic capacity of the bottleneck increased and there were no severe competitions in front of the bottleneck. While inside the bottleneck, the density reached 7.7 ped/m$^2$ because of the limited area.

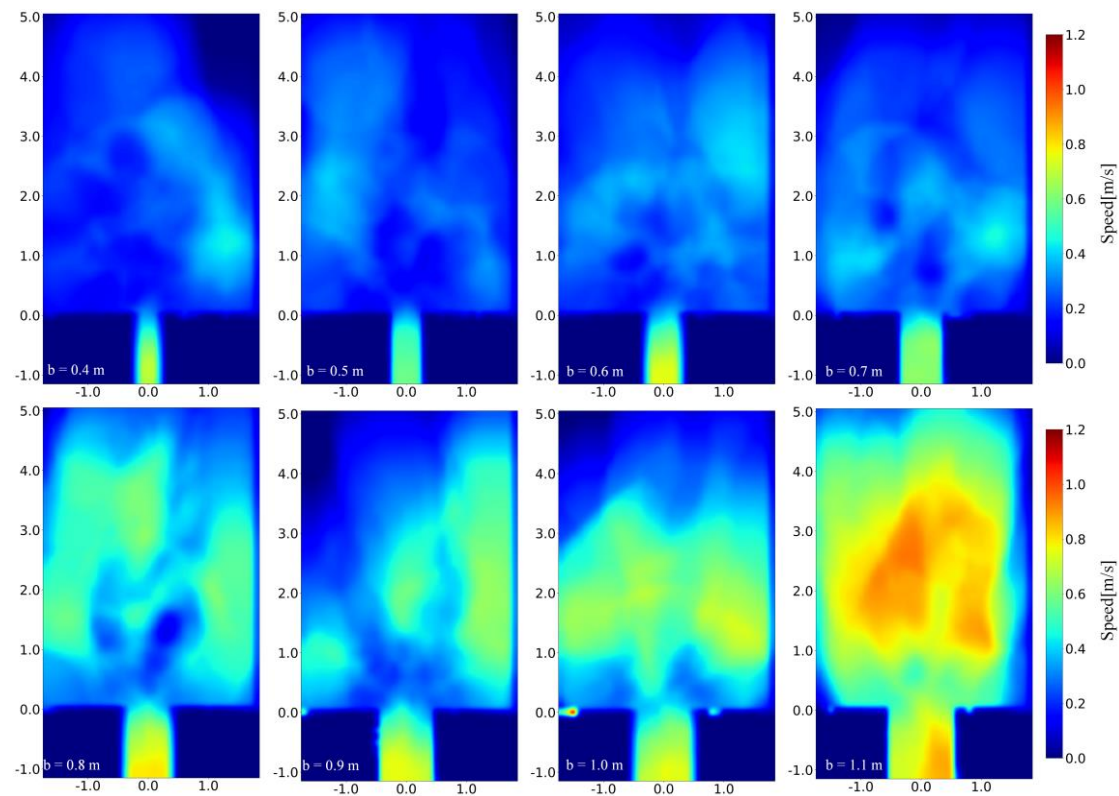

Fig. 7 Speed profiles for bottleneck of varying widths. The speeds in front of the bottleneck increase with the increasing bottleneck width, while the lowest speed appears around the bottleneck. The speeds inside the bottleneck show no obvious discrepancies for different bottleneck width and they oscillate around 1.0 m/s.

We further study the distribution of speed around the bottleneck in Fig. 7. Children passed through the bottleneck with less deceleration with the increase of the bottleneck width. The speed of crowds in front of the bottleneck is obviously smaller than that in other regions. It increases from 0.2 m/s to about 0.6 m/s with the increase of the bottleneck width. Besides, children increase their speed gradually when they are leaving the bottleneck. Despite of the high speed over 2.5 m/s at the beginning of each run (see Fig. 2), the mean speed of the kid is surprisingly around 1.0 m/s inside the bottleneck nearly for all the widths, which is comparable with that of the adults walking in normal speed (see Fig. 8 (b)) (Liddle et al., 2009). Even with a higher motivation, the frontal pedestrians, restricted area and fierce competition in front of the bottleneck make the children fail to run at a high speed inside the bottleneck. As a result, the speed distributions inside the bottleneck are similar in spite of the different motivations in the experiments with children and adults (Liddle et al., 2011b; Seyfried et al., 2009).

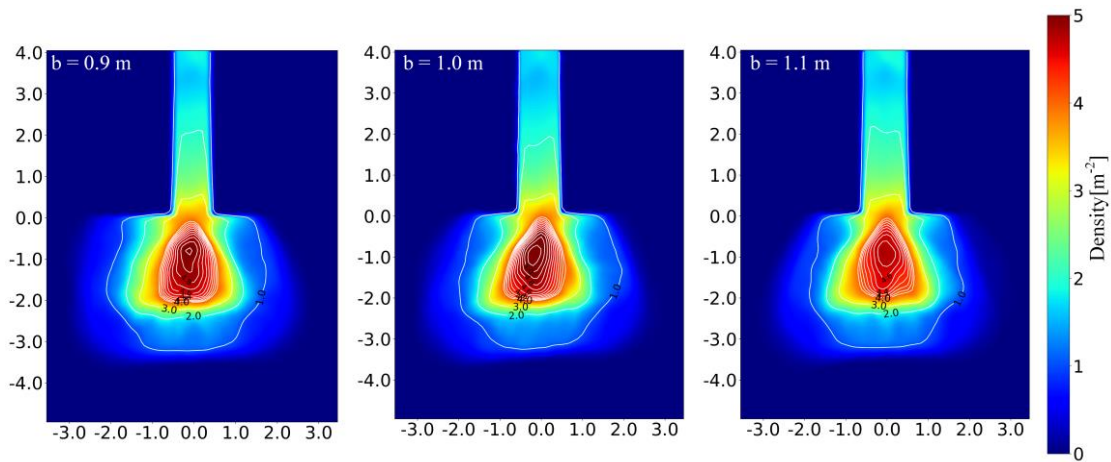

(a) Density profiles

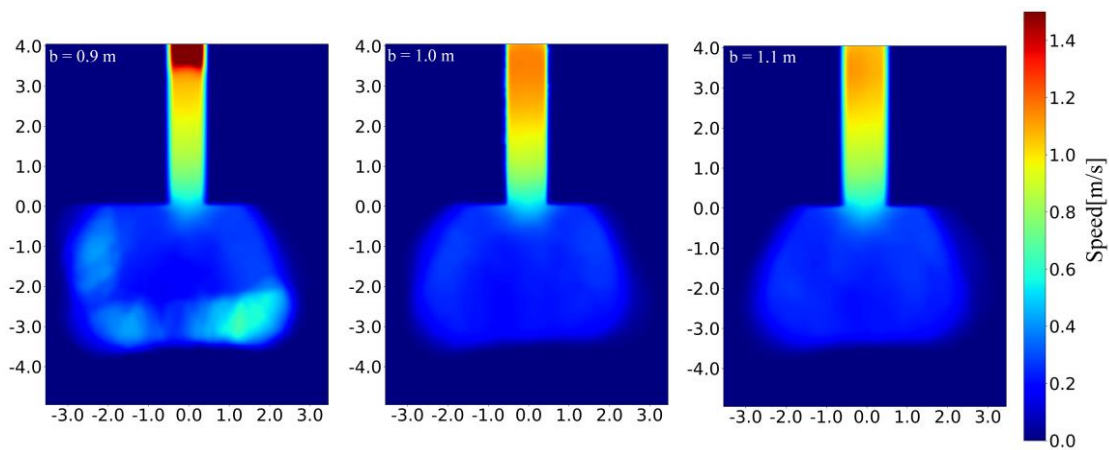

(b) Speed profiles

Fig. 8 Density and speed distribution for the adults moving through different wide bottleneck in Ref (Liddle et al., 2009). The congested states are from 120 frame to 2050 frame at b = 0.9 m, from 350 frame to 2100 frame at b =

0.9 m, from 300 frame to 2000 frame at b = 1.1 m. The density distribution is teardrop-shaped in front of the bottleneck and the peak density is about 5 ped/m$^2$ which lies around 100 cm away from the bottleneck entrance. The speed inside the bottleneck oscillates around 1.0 m/s despite the width, which is similar to children movement.

3.3 Flow

In this section, we study the relation between the flow $J$ and bottleneck width $b$. Here, flow is defined as the number of pedestrians who enters the bottleneck (y = 0.0 m) per unit of time.

In this analysis, only the data of the first 30 children are considered to calculate the flow using in each run. As shown in Fig. 9 left, the flow shows an upward trend with the increasing width, which agrees well with previous studies. We further make a linear regression on data and it follows $J = 5.11b - 0.95$. The result of independent-sample test is p= 5.3E-5 and R$^2$=0.94, which indicates that the flow shows significant dependency on the bottleneck width. Compared with that in the previous studies (Kretz et al., 2006; Liddle et al., 2009; Müller, 1981; Nagai et al., 2006; Seyfried et al., 2009), however, the slope of the relation in this study is significantly larger as can be seen from the subgraph of Fig. 9 left. The flow of children increases faster with the increase of the bottleneck width than that of adults. It is worth mentioning that the participants of those previous experiments were adults and walked in normal speed. When only the number of the participants is considered in flow calculation, these two main differences may lead to the different flows. A given bottleneck can accommodate more children than adults. Different motivations in these experiments result in different free speeds and behaviors like pushing or cooperation among pedestrians.

In order to exclude the influence of participants' physical size and motivation, we compare the data with two previous experiments in (Liddle et al., 2009) and (Seyfried et al., 2009) for similar bottleneck geometry and available complete trajectory data by defining two new variables $b^*$ and $J^*$. The flow follows $J = 2.03b - 0.29$ (Independent-sample test: p=4.14E-7, R$^2$=0.96) and $J = 2.60b - 0.73$ (Independent-sample test: p=7.37E-4, R$^2$=0.98) in (Liddle et al., 2009) and (Seyfried et al., 2009) respectively. The definition of $b^*$ and $J^*$ can be found in Equation 1 and 2. Here, we qualify the motivation with the speed inside the bottleneck. Despite of the higher speed of the children at the beginning of each run, the speed inside the bottleneck is around 1.0 m/s which is similar to adults (Liddle et al., 2011a) (see Fig. 7 and Fig. 8 (b)). Hence, we could suppose that the influence of the movement motivations of the children on the outflow of the bottleneck is similar to that of adults. Under these assumptions, we plot the relation between $J^*$ and $b^*$ in Fig 9 right. With such transformation, the proposed variable $J^*$ follows $J^* = 0.20b^* - 0.13$ (Independent-sample test: p = 5E-5, R$^2$=0.94) in this study, While they are $J^* = 0.21b^* - 0.062$ (Independent-sample test: p = 4.31E-7, R$^2$=0.96) and $J^* = 0.27b^* - 0.16$ (Independent-sample test: p = 6.92E-4, R$^2$=0.98) in (Liddle et al., 2009) and (Seyfried et al., 2009) respectively. Surprisingly, the relations between $J^*$ and $b^*$ agree well for the different experiments. This implies that the dependency between flow and bottleneck width can be unified for different pedestrians and movement motivations when taking out of the influence of body size and motivation.

$$b^* = b/shoulder\_breadth \qquad (1)$$
$$J^* = J \times chest\_depth \ /motivation \qquad (2)$$

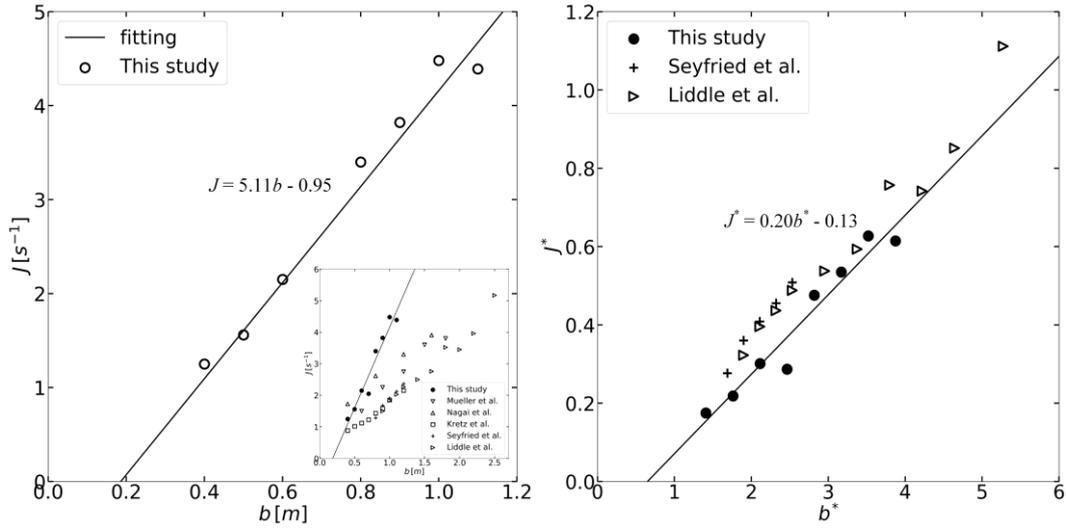

Fig. 9 Relation between the flow and bottleneck width. Left: the results obtained from the experiment of children. Note that only the first 30 children are considered in the flow calculation. The subfigure shows the comparison with the relations from the previous experiment with adults and obvious differences can be observed especially for the slope of the data. Right: The unified relation between $b^*$ and $J^*$ by considering pedestrian's body size and movement motivation.

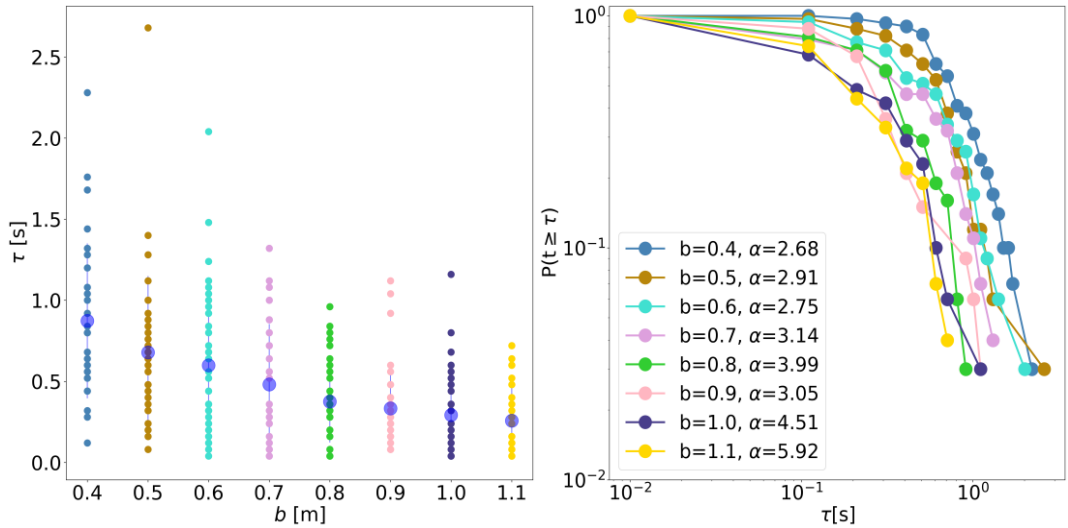

Fig. 10 The range of the time intervals $\tau$ (left) and the complementary cumulative probability of time intervals $\tau$ (right). The data of the first 5 children, the last 10 children in each run and the uncooperative children were removed to avoid the transient state. $p(\tau)$ is consistent with the existence of a power-law tail: $p(\tau) \sim \tau^{-\alpha}$. The corresponding power-law exponent $\alpha$ increases with the increase of the bottleneck. Higher value of $\alpha$ indicates the presence of shorter time intervals between two consecutive pedestrians and smaller probability of having longer clogs.

The time interval $\tau$ of two consecutive children for all experiments of different widths are studied and the complementary cumulative probability distribution function (CCDF, survival function) of the time intervals are presented in Fig. 10. Here the time interval is defined as the time elapsed between two consecutive pedestrians passing the bottleneck. From the recorded videos, it is clear that children can pass the bottleneck without congestion during the first few seconds and the last few seconds. Consequently, we remove the first 5 children and the last 10 children in each run to avoid the impact of the transient state. Besides, the data of some children in b=0.4 m and b=0.7 m are also removed in the analysis, since these children behaved uncooperatively and obstructed the exit and influenced others behavior. As shown in Fig. 10 left, time interval shows a downward trend with the increase of the bottleneck width. Similar to (Pastor et al., 2015), $p(\tau)$ is consistent with the existence of a power-law tail: $p(\tau) \sim \tau^{-\alpha}$ as shown in Fig. 10 right. The fitting results are obtained based on the Clauset-Shalizi-Newman method (Clauset et al., 2009). In similar cases, the exponent $\alpha$ can measure the dynamics of the egress (Lin et al., 2016; Pastor et al., 2015). It suggests that the slopes of the curves exhibit a strong dependency on the bottleneck width. Narrower bottleneck exhibits a longer tail, which means the greater possibility of having longer clogs. In this study, the corresponding power-law exponent $\alpha$ is the highest for the widest bottleneck. The value of $\alpha$ presents rising trend with the increase of the bottleneck width on the whole with two exceptions (b=0.6 m and b=0.9 m). By checking the recorded videos, we observe that (1) when b = 0.6 m, a kid slowed down inside the bottleneck deliberately for about 1 second which affected the movement of crowds slightly; (2) when b = 0.9 m, the bottleneck can accommodate 3 to 4 children, while for b = 0.8 m it can only hold 2 to 3 children. Compared with 0.8 m wide bottleneck, the competition among children around 0.9 m wide bottleneck is more intense (see Fig 5 and Fig 6) which reduces the outflow through the bottleneck. An increase of the bottleneck width results in a higher value of $\alpha$, indicating the presence of shorter time intervals between the passages of two consecutive pedestrians.

## 4. Summary

In this study, an experiment on the movement of children through bottlenecks was performed under controlled conditions to narrow the gap of research on children's evacuation characteristics. In total, about 50 children aged 3-5 years old participated in the experiment. They were asked to pass the bottleneck as quickly as possible before the experiment. We mainly focus on the impact of varying bottleneck widths (0.4 m to 1.1 m at 0.1 m intervals) on the density, speed and flow of the crowds. Several interesting phenomena are observed from the recorded videos. The children responses differently to the same instruction and behaviors that might reduce evacuation efficiency are observed. Despite of the higher speed over 2.5 m/s at the beginning of each run, the speed of children is about 1.0 m/s inside the bottleneck, which is similar to the speed of adults in previous studies (Liddle et al., 2011a). Besides, compared with tear-drop shape density distribution of adults in normal condition (Liddle et al., 2011a), arch-like blockings of the exit appear in this experiment since children compete against each other fiercely. The density of children in front of the bottleneck (up to 14 ped/m$^2$) is obviously higher than that of adults for smaller body size and higher motivations. With the increase of the bottleneck width, the density decreases gradually for enlarged traffic capacity. The locations of the peak density lie around 30 cm from the bottleneck entrance in this study (*b* ranges from 0.4 m to 1.0 m), whereas they are around 100 cm from the entrance of the bottleneck (*b* = 0.9, 1.0, 1.1 m) in the former study (Liddle et al., 2011a)

Moreover, the flow shows a growth trend with the increase of the bottleneck width, which is similar to the relation from adults' experiments previously (Kretz et al., 2006; Liddle et al., 2009; Müller, 1981; Nagai et al., 2006; Seyfried et al., 2009). However, the values are significantly higher in this study for the same bottleneck width. When excluding the impact of participants' body size and movement motivation, similar relation of flow and bottleneck width between running children and walking adults is found. Besides, it is found that the probability of the time interval $\tau$ follows the existence of a power-law tail: $p(\tau) \sim \tau^{-\alpha}$ and the value of $\alpha$ increases with the increase of the bottleneck width, which indicates the greater possibility of having longer clogs at narrower bottleneck.

The findings in this study suggests that it is necessary to conduct evacuation drills in buildings for children and to teach them how to escape from emergency. Further studies should be carried out to study the children's movement characteristics through different building structures.

# Acknowledgements

This work was supported by the National Natural Science Foundation of China (Grant No. 71704168); the Anhui Provincial Natural Science Foundation (Grant No. 1808085MG217); the Fundamental Research Funds for the Central Universities (Grant No. WK2320000040); the Fundamental Research Funds for the Central Universities (Grant No. WK2320000040); the State Key Laboratory of Fire Science in University of Science and Technology of China (Grant No. HZ2018-KF12); CNTC Hubei Provincial Tobacco Corporation "YOUNG ELITE SCIENTISTS SPONSORSHIP PROGRAM" (Grant No. 027Y2018-036) and the grant from the Research Grants Council of the Hong Kong Special Administrative Region (Project No. CityU 11301015).